\documentclass[epj,draft]{svjour}
\usepackage{pstricks,bbm}
\usepackage{psfig}

\begin{document}
\title{Magnetism and Phase Separation in the Ground State of the Hubbard Model}
\author{R.~Zitzler\inst{1} \and Th.~Pruschke\inst{2} \and R.~Bulla\inst{2}}
\authorrunning{R.~Zitzler et al.}
\titlerunning{Magnetism and Phase Separation $\ldots$}
\offprints{R.~Zitzler}
\institute{Institut f\"ur Theoretische Physik I, Universit\"at
Regensburg, 93040 Regensburg, Germany
\and Theoretische Physik III,  Elektronische Korrelationen und
Magnetismus, Institut f\"ur Physik, Universit\"at Augsburg, 86135 Augsburg,
Germany}
\date{Received: date / Revised version: date}
%
\abstract{We discuss the ground state magnetic phase diagram of the Hubbard
model off half filling within the dynamical mean-field theory. The effective
single-impurity Anderson model is solved by Wilson's numerical renormalization
group calculations, adapted to symmetry broken phases. We find
a phase separated, antiferromagnetic state up to a critical doping for small
and intermediate values of $U$, but could not stabilise a N\'eel state 
for large $U$ and finite doping. At very large $U$, the phase diagram
exhibits an island with a
ferromagnetic ground state. Spectral properties in the ordered phases are
discussed.
}
%
\maketitle
\section{Introduction\label{sec:intro}}
Originally proposed for the description of ferromagnetism in transition metals,
the Hubbard model \cite{hgk}
\begin{equation}\label{equ:HM}
H=\sum\limits_{i,j}\sum\limits_\sigma t_{ij}c^\dagger_{i\sigma}c^{\phantom{\dagger}}_{j\sigma}
+U\sum\limits_i n_{i\uparrow}n_{i\downarrow}
\end{equation}
is the simplest model to describe the interplay between delocalization
or band formation in solids and the local Coulomb correlations. Despite its
simplicity, the phase diagram of the Hubbard model (\ref{equ:HM}) reveals a surprising richness. One finds
 Mott-Hubbard type metal-insulator transitions \cite{mit}, antiferromagnetism
\cite{afm}, ferromagnetism \cite{fm} and incommensurate magnetic phases \cite{incomm}.
More recently, the Hubbard model has also become one of the most promising 
candidates
to describe the low-energy properties and possibly the superconductivity in the
high-$T_c$ cuprates \cite{hightc}.

The Hubbard model at half filling $\langle n\rangle=1$ has been
investigated thoroughly using
various approximate and exact techniques \cite{hub_pre,hub_rigo} and its
properties are understood to a large extent. Off half filling, the model is
well understood  for $d=1$ but the situation is less clear in dimensions
$d>1$.
Basically the only rigorous result is due to Nagaoka \cite{nagaoka}, who
proved that a ferromagnetic ground state is possible under
certain conditions.

The introduction of the limit $D\to\infty$ \cite{mv89} in principle allows
to solve models like the Hubbard model exactly without
loosing the competition between kinetic energy and local
Coulomb repulsion 
\cite{review_dmft}. This surprising insight has subsequently triggered a large
amount of investigations of the infinite dimensional
 Hubbard model  \cite{review_dmft}.
In addition, $D\to\infty$ turned out
to be a reasonable starting point for weak-coupling expansions
\cite{PvD_wk,PvD_ps}. Within this approach, in addition to the expected
magnetic order for a bipartite lattice, phase separation was found for the
whole region of the magnetically ordered phase \cite{PvD_ps}. Since this
result is based on a weak-coupling expansion, it is far from clear whether
it holds for finite values of the interaction $U$ as well. Results from a
numerically exact solution of the Hubbard model in $D=\infty$ based on Quantum
Monte-Carlo simulations for example showed no evidence for phase separation
\cite{jim_mark}, but these calculations were done in the paramagnetic phase
and at finite, comparatively high temperatures.

The question whether phase separation in the Hubbard model occurs in a
certain parameter regime is of some importance for two reasons. First,
from a model theoretical point of view, it is of course interesting to
explore the stability of the different possible ordered phases 
which might be unstable with respect to
phase separation. Second, a vicinity to
phase separation has been discussed as one of the possible ingredients to the superconductivity in the high-$T_c$ cuprates \cite{kivelson,castellani}.
Moreover, a tendency towards phase separation together with the long-range part of
the Coulomb interaction may in principle lead to charge ordered states 
such as stripe-phases.

Phase separation has long been predicted \cite{kivelson} and indeed been
observed for the $t$-$J$ model in  $D=1,\,2$ \cite{dagotto}. Since the
$t$-$J$ model for vanishing $J$ is connected to the Hubbard model in the
limit $U/t\to\infty$ \cite{fulde}, additional information about phase
separation in the strong coupling limit could thus be obtained. 
 The critical $J$, below which phase separation vanishes in the $t$-$J$
model is typically of the order of $t$ \cite{dagotto}; this seems to
rule out phase separation in the Hubbard model at least for $U/t\to\infty$
in $D=2$. However, the situation has not been clarified yet, 
because direct
inspection of the Hubbard model in the limit $U/t\to\infty$ leads to contradictory
results \cite{su,Tandon99}.

The results for the $2D$ Hubbard model for finite $U$ available 
so far have not
revealed signs for phase separation \cite{dagotto,becca}. However, these
results are typically based on Quantum Monte-Carlo or related techniques, which
have severe  problems in the interesting parameter regime close to
half filling and at very low temperatures. Consequently, one either has to
restrict oneself to rather small system sizes \cite{dagotto} or use further
approximations \cite{becca}.
Thus, to our present knowledge a detailed study of the ground state phase
diagram of the Hubbard model in the thermodynamic limit and in the vicinity
of half filling, comprising weak, intermediate and strong coupling within a
non-perturbative approach is not available.

Such an approach is provided by the limit $D\to\infty$, which allows for
in principle exact calculations in the thermodynamic limit for all model
parameters, even at $T=0$. The price one pays is the neglect of non-local
dynamics, which of course is most severe for $D\le2$. Nevertheless, the theory
can give valuable information about whether phase separation is possible
at all. 

The paper is organized as follows.
In section \ref{sec:method}, we  give a brief description of the solution of the
Hubbard model in the limit $D\to\infty$. 
Results for the phase diagram and the dynamics in the different
phases are presented in section \ref{sec:results}. 
The main conclusions of the paper are summarized in section \ref{sec:summary}.

\section{Theoretical background\label{sec:method}}
\subsection{General remarks}
The dynamical mean-field theory (DMFT) to exactly solve
the Hubbard model in the limit $D\to\infty$ is based on the work by Metzner
and Vollhardt \cite{mv89} and is by now well-established \cite{review_dmft}.
The basic ingredient is that for $D\to\infty$ the proper single-particle
self energy $\Sigma(\vec{k},z)$ becomes purely local or momentum independent,
i.e.\ $\Sigma(\vec{k},z)\stackrel{D\to\infty}{\longrightarrow}\Sigma(z)$
\cite{mv89,muha89}.
This can be used to map the Hubbard model (\ref{equ:HM}) onto
an equivalent quantum impurity problem supplemented by a self-consistency
condition \cite{review_dmft}. The remaining problem (the solution of a quantum
impurity model) is, however, highly nontrivial. Several
approximate and numerically exact techniques are currently 
available \cite{review_dmft,nrg}.

Most of these methods cannot access $T\to0$ or are restricted to
the weak-coupling regime of the Hubbard model. The most reliable technique
to solve the quantum impurity problem for all interaction strenghts
$U$ and fillings $n$ at $T=0$ and low $T$ is the numerical
renormalization group (NRG) \cite{nrg,krish}. Originally, this method was set
up to treat the paramagnetic problem only \cite{krish}, but recent extensions
have shown that calculations with a symmetry breaking field are possible
with a similar level of accuracy, too
\cite{nrg_magn_costi,nrg_magn_walter}. Hence  we are able to study
magnetically ordered phases directly at $T=0$.

In contrast to the standard NRG, a more refined approach
has to be used to calculate dynamical quantities in the presence of a magnetic
field. This has first been noted
by Hofstetter, who observed discrepancies in the magnetization calculated
from the spectral functions and the ground state occupation numbers
\cite{nrg_magn_walter}. To resolve this problem, he proposed a modification
of the standard method \cite{sakai} to calculate the spectral function. A more
detailed discussion of this technical point and its physical background will
be presented elsewhere.

There are in principle two ways to determine the phase boundary between the paramagnetic
and a magnetically ordered state. First, one can calculate the susceptibility
corresponding to the anticipated order and look for a divergence. Second, one
can allow for a proper symmetry breaking in the one-particle Green function
and search for the region in parameter space where a solution with broken
symmetry becomes stable. Especially for $T=0$ the first method is rather
cumbersome in general and, by construction, also makes no statement about the
thermodynamic stability of phases beyond the critical point.

We thus use the second approach as our method of choice. However, this prohibits the search
for incommensurate phases, because only broken symmetries with a commensurate
wave vector can be implemented that way. Since we are interested mainly in
standard N\'eel type antiferromagnetic order, the proper way is to introduce
an AB-lattice structure and allow for different sublattice magnetizations. The
resulting Green function then becomes a $2\times2$ matrix, which within the
DMFT has the form
\cite{review_dmft}
\begin{equation}
  \label{equ:Gaf}
  \tens{G}_{\vec k\sigma}(z)=\left(\begin{array}{c@{\ \ }c}
z+\mu-\Sigma^{\rm A}_\sigma(z) & -\epsilon_{\vec k}\\
-\epsilon_{\vec k} & z+\mu-\Sigma^{\rm B}_\sigma(z) 
\end{array}\right)^{-1}\;\;,
\end{equation}
where $\vec k$ is a vector in the magnetic Brillouin zone (MBZ). For the N\'eel
state on an AB-lattice a further simplification arises from the symmetry
$\Sigma^{\rm A}_\sigma(z)=\Sigma^{\rm B}_{\bar{\sigma}}(z)$. For the
calculation this means that we do not have to solve independent quantum
impurity models for the two sublattices, but only one for say sublattice A.

\begin{figure}[htb]
%
%
\begin{center}
\unitlength1truemm
\normalbaselineskip 11pt
\normalbaselines
\begin{picture}(80,61)
\thicklines
\put(0.00,50.00){\framebox(80,10.00)[lc]
{
\parbox{75mm}{ Choose initial chemical potential
$\mu$ and self-energy $\Sigma(z)$}}}

\put(0.00,0.00){\framebox(80.00,50.00)[cc]{}}

\put(0.00,45.00){
\parbox{70mm}{ Calculate effective medium $\Gamma(\omega)$}}

\put(10.0,7.5){\framebox(70.00,32.5)[cc]{}}

\put(20.00,30.00){\framebox(60.00,10.00)[cc]
{\parbox{55mm}{ Solve effective SIAM defined by $\mu$, $U$
and $\Gamma(\omega)$}}}

\put(20.00, 20.00){\framebox(60.00,10.00)[cc]
{\parbox{55mm}{ Determine $\langle n\rangle_{\rm SIAM}$ from solution}}}

\put(10.00, 12.5){{
\parbox{65mm}{ Adjust $\mu$ to obtain desired $\langle n\rangle$}}}

\put(0.00,2.0){
\parbox{70mm}{ Iterate to self-consistency of $\Sigma(z)$}}
\end{picture}
\unitlength1pt
\end{center}
\caption{Flow diagram for the DMFT self-consistency cycle with fixed filling $\langle n\rangle$.} 
\label{fig:flow} 
\end{figure}
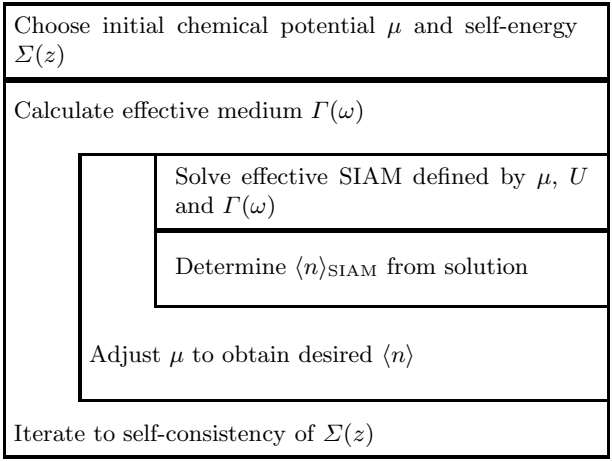

But how to control the filling if the homogeneous solution 
(homogeneous concerning the charge distribution) turns out
to be unstable towards phase separation? Fixing the chemical potential
$\mu$ is not sufficient as the system will be driven to a filling corresponding
to a stable solution, such as $n=1$. To enforce a metastable state with
finite doping we adopt a procedure which has been already used in 
calculations for the half-filled Hubbard model in a homogenous magnetic field 
\cite{laloux}. The schematic flow
diagram of the resulting DMFT self-consistency cycle is shown in Fig.~\ref{fig:flow}. Starting from a
paramagnetic solution for the desired doping, a homogenous or staggered
symmetry breaking is introduced and the corresponding effective medium,
$\Gamma_\sigma(\omega)$, for the DMFT cycle is determined \cite{review_dmft}. 
Keeping the medium $\Gamma_\sigma(\omega)$ fixed, one now varies the on-site
energy of the effective SIAM until the desired doping has been reached. This
result is used to obtain a new effective medium, and the procedure is repeated
until convergency is reached. It should be noted that, for a metastable
state, no true convergency can be reached in the sense that the solution,
when iterated further without adjusting the filling properly, will flow again
into the phase separated one. Typically, 
for a converged calculation, the on-site energy between successive DMFT
iterations shows a weakly damped oscillatory behavior but does not change by
more than 2-3\%. In order to minimize errors introduced by these oscillations
we average all quantities over several iterations.

To find the correct ground state, we need to calculate the ground state energy
\begin{equation}\label{equ:E}
\frac{E}{N} = \frac{1}{N}\langle H\rangle = \frac{1}{N}\langle H_t\rangle + 
\frac{U}{N}\sum_{i}\langle n_{i\uparrow}n_{i\downarrow}\rangle\;\;,
\end{equation}
where $H_t$ is the kinetic part of the Hamiltonian (\ref{equ:HM}). The
expectation value $\langle n_{i\uparrow}n_{i\downarrow}\rangle$ can be
determined within the NRG directly. The quantity $\langle H_t\rangle$, on the
other hand, depends on the phase we are looking at. For the para- and ferromagnetic
phases it is simply given by \cite{review_dmft}
\begin{equation}
  \label{equ:Htpm}
  \frac{1}{N}\langle H_t\rangle=\sum\limits_\sigma\int\limits_{-\infty}^\infty 
d\epsilon\ 
\epsilon\ \rho^{(0)}(\epsilon)
\int\limits_{-\infty}^\infty d\omega f(\omega) A_\sigma(\epsilon,\omega)\;\;,
\end{equation}
with $\rho^{(0)}(\epsilon)$ the density of states (DOS) for the non-inter\-ac\-ting
system, $f(\omega)$ the Fermi function and
$$
A_\sigma(\epsilon,\omega)=-\frac{1}{\pi}\Im m\frac{1}{\omega+\mu-\epsilon-\Sigma_\sigma(\omega+i0^+)}
$$
the spectral function of the Hubbard model in the DMFT, i.e.\ with $\vec k$-independent
one-particle self energy.

For an antiferromagnetic state with N\'eel
order one has to take into account the AB-lattice structure and the
formula becomes \cite{review_dmft}
\begin{equation}
  \label{equ:Htafm}
  \frac{1}{N}\langle H_t\rangle=2\int\limits_{-\infty}^\infty 
d\epsilon\ 
\epsilon\ \rho^{(0)}(\epsilon)
\int\limits_{-\infty}^\infty d\omega f(\omega) B(\epsilon,\omega)
\end{equation}
instead, with
$$
B(\epsilon,\omega)=-\frac{1}{\pi}\Im m
\frac{1}{\sqrt{\zeta_\sigma(\omega)\zeta_{\bar{\sigma}}(\omega)}-\epsilon}
$$
and $\zeta_\sigma(\omega)=\omega+\mu-\Sigma_\sigma(\omega+i0^+)$. 
Obviously,
expression (\ref{equ:Htafm}) reduces to (\ref{equ:Htpm}) without magnetic order, i.e.\ $\zeta_\sigma(\omega)=\zeta_{\bar{\sigma}}(\omega)$.

Throughout the rest of the paper we concentrate on results for a simple
hypercubic lattice and
$$
t_{ij}=\left\{\begin{array}{l@{\ \ \ }l}
-t & \mbox{if $i$, $j$ are nearest neighbors}\\
0 & \mbox{else}\end{array}\right.\;\;.
$$
The resulting DOS $\rho^{(0)}(\epsilon)$ then becomes
\cite{mv89,muha89}
\begin{equation}
  \label{equ:DOS}
  \rho^{(0)}(\epsilon)=\frac{1}{t^*\sqrt{\pi}}e^{-(\epsilon/t^*)^2}\;\;.
\end{equation}
In the following we use $t^*=2\sqrt{D}=1$ as our unit of energy.

\subsection{Weak-coupling results}

Let us briefly review some  weak-coupling results as these will
be frequently referred to in section
\ref{sec:results}.
Since the hypercubic lattice is a bipartite lattice, one obtains in lowest
order, i.e.\ in Hartree approximation, a transition into a N\'eel state
for any $U>0$ at $T=0$ below a critical doping $\delta_c^{\rm H}(U)$. For small
$U\to0$ the
magnetization $m$ as well as the critical doping depend non-analytically on U,
i.e.\ $m,\delta_c^{\rm H}\propto\exp\left(-1/(U\rho^{(0)}(0))\right)/U$ independent
of the dimension.

A quantity of particular interest in the DMFT is the single-particle Green
function.  The general structure of the Green function in the N\'eel
phase for both  Hartree theory and DMFT is  given by expression
(\ref{equ:Gaf}),
where in the Hartree approximation $\Sigma_\sigma(z)$ reduces to
$\Sigma_\sigma^{\rm H}(z)=Un_{\bar{\sigma}}=\frac{1}{2}U(n-\sigma m)$
with $n$ the filling and $m$ the magnetization.
The local Green function is obtained from (\ref{equ:Gaf}) by summing over
$\vec k\in{\rm MBZ}$, which yields for example for spin up
\begin{equation}
  \label{equ:Gafup}
  G_\uparrow(\omega)=\frac{\zeta_\downarrow(\omega)}{%
\sqrt{\zeta_\uparrow(\omega)\zeta_\downarrow(\omega)}}
\ G^{(0)} \!
\left(\sqrt{\zeta_\uparrow(\omega)\zeta_\downarrow(\omega)}\right)
\end{equation}
with $\zeta_\sigma(\omega)=\omega+i0^++\mu-\frac{U}{2}n+\sigma\frac{U}{2}m$
and 
\begin{equation}\label{equ:G0}
G^{(0)}(z)=\int\limits_{-\infty}^\infty d\epsilon\; \frac{\rho^{(0)}(\epsilon)}{z-\epsilon}\;\;.
\end{equation}
For the further discussion let us define
$$
\begin{array}{l@{\ = \ }l}
\displaystyle\omega_- & \frac{U}{2}n-\mu-\frac{U}{2}m\\[5mm]
\displaystyle\omega_+ & \frac{U}{2}n-\mu+\frac{U}{2}m
\end{array}
$$
Then, as long as $\omega\le\omega_-$ or $\omega\ge\omega_+$, the radicant in
(\ref{equ:Gafup}) is positive and the resulting DOS can be expressed as
$$
\rho_\uparrow(\omega)=\frac{\zeta_\downarrow(\omega)}{%
\sqrt{\zeta_\uparrow(\omega)\zeta_\downarrow(\omega)}}\rho^{(0)}
\left(\sqrt{\zeta_\uparrow(\omega)\zeta_\downarrow(\omega)}\right)\;\;.
$$
For $\omega_i<\omega<\omega_+$, on the other hand, the radicant in (\ref{equ:Gafup})
is negative, i.e.\ $\sqrt{\zeta_\uparrow(\omega)\zeta_\downarrow(\omega)}=
i\sqrt{|\zeta_\uparrow(\omega)\zeta_\downarrow(\omega)|}$. Since for the
particle-hole symmetric DOS (\ref{equ:DOS}) the Green function $G^{(0)}(z)$
defined in (\ref{equ:G0}) for purely
imaginary arguments is purely imaginary, too, one finds
$$
\rho_\uparrow(\omega)=0\;\;,
$$
i.e.\ the DOS has a gap between $\omega_-$ and $\omega_+$.
\begin{figure}[htb]
\begin{center}
\mbox{}
\psfig{figure=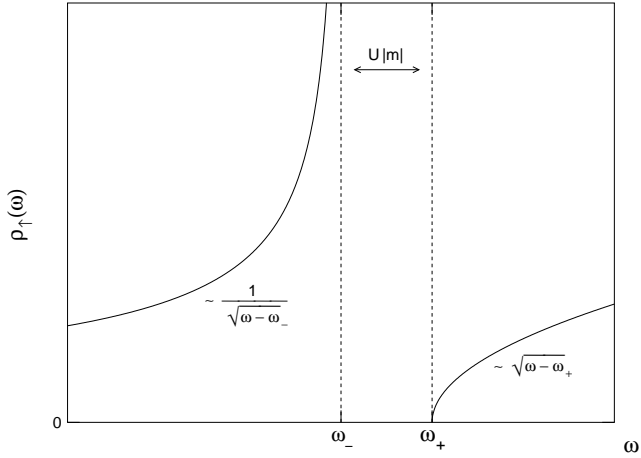,width=0.47\textwidth}
\end{center}
\caption{Behavior of the DOS in Hartree approximation close to the gap edges.%
\label{fig:DOSH}}
\end{figure}
As one approaches $\omega_-$ from below or $\omega_+$ from above, it is easy
to confirm that
\begin{equation}
\label{equ:rhoGap}
\rho_\uparrow(\omega)\approx\left\{\begin{array}{l@{\ \ \ }l}
\displaystyle\sqrt{\frac{Um}{|\omega-\omega_-|}}\rho^{(0)}(0)
&
\displaystyle\omega\nearrow\omega_-\\[5mm]
\displaystyle\sqrt{\frac{|\omega-\omega_+|}{Um}}\rho^{(0)}(0)
&
\displaystyle\omega\searrow\omega+
\end{array}\right.\;\;\;.
\end{equation}
The corresponding DOS for $\sigma=\downarrow$ has a similar behavior. Here,
however, the DOS diverges like $1/\sqrt{|\omega-\omega_+|}$ at the upper gap
edge, and vanishes like $\sqrt{|\omega-\omega_-|}$ at the lower one.

In order to determine the thermodynamically stable phase
one has to calculate the ground state energy as function of the doping
$\delta=1-n$. The result up to second order in $U$ is \cite{PvD_ps}
\begin{equation}
  \label{equ:Egwk}
  E(\delta)-E(0)=-\frac{U}{2}\delta+\alpha^{\rm H}\cdot\Phi^{\rm H}(\delta/\delta_1)\;\;,
\end{equation}
where
\begin{equation}
  \label{equ:Phi}
  \Phi^{\rm H}(x)=\left\{\begin{array}{l@{\ \ }l}
  \displaystyle\frac{1}{2}x\left(1-\frac{1}{4}x\right) & x<1\\[5mm]
  \displaystyle\frac{1}{4}\left(1+\frac{1}{2}x^2\right) & x>1
  \end{array}\right.
\end{equation}
and $\delta_1$ is the critical doping for antiferromagnetism in Hartree
approximation. The coefficient
$\alpha^{\rm H}$ is given by $\alpha^{\rm H}=2\delta_1^2/\rho^{(0)}(0)$.
\begin{figure}[htb]
\begin{center}\mbox{}
\psfig{figure=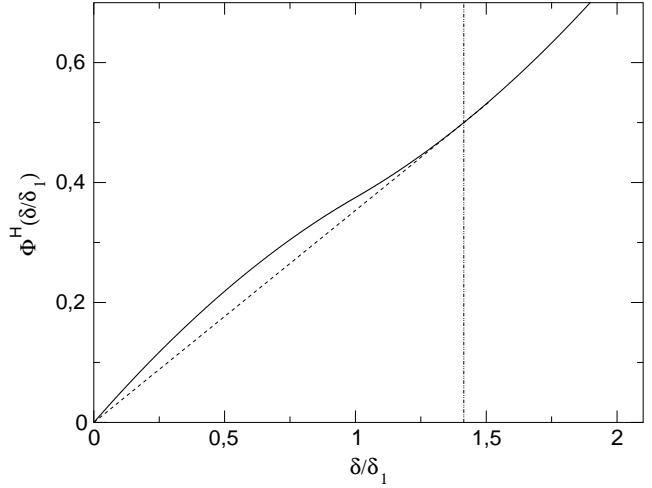,width=0.47\textwidth}
\end{center}
\caption{The function $\Phi^{\rm H}(\delta/\delta_1)$ from Eq.~(\ref{equ:Phi}). Note the
concave curvature between $\delta=0$ and $\delta=\sqrt{2}\delta_1$. The
dashed line shows the actual behavior of the ground state energy following
from a Maxwell construction.\label{fig:Phiwk}}
\end{figure}
The function $\Phi^{\rm H}(\delta/\delta_1)$ appearing in expression (\ref{equ:Egwk})
leads to the full line in Fig.~\ref{fig:Phiwk}. Apparently, this function is
not convex for small $\delta$, i.e.\ the resulting phase is thermodynamically
unstable towards phase separation for dopings less than $\delta_c=\sqrt{2}\delta_1$.
The resulting ground state energy is then obtained from a Maxwell construction,
given by the straight dashed line in Fig.~\ref{fig:Phiwk}.

\section{Results\label{sec:results}}
Let us start with a short
\begin{figure}[htb]
\begin{center}
\mbox{}
\psfig{figure=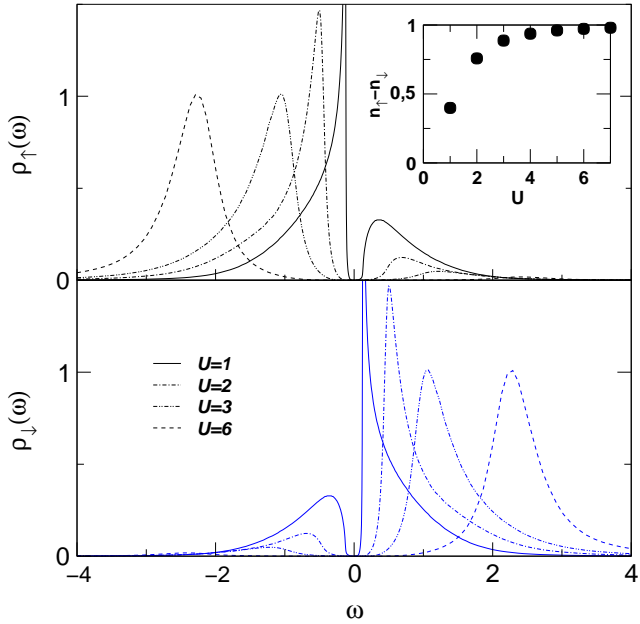,width=0.47\textwidth}
\end{center}
\caption{DOS for spin up and down at half filling in the antiferromagnetic
phase as function of $U$. While for small values of $U$ the weak-coupling
form (\ref{equ:rhoGap}) is approximately reproduced, the DOS for large $U$
is basically that of the Mott-Hubbard insulator. The inset shows the magnetization
as function of $U$.\label{fig:DOShalf}}
\end{figure}
overview of the behaviour at half filling, $n=1$. Here, the N\'eel phase is
energetically stable. The variation of the DOS for increasing $U$ from
$U=1$ (full curve) to $U=6$ (dashed curve) is shown in Fig.~\ref{fig:DOShalf}.
As expected, the DOS for small $U$ resembles the form (\ref{equ:rhoGap})
predicted for weak-coupling, i.e.\ one sees the remnants of the characteristic
square-root divergency in the spin up DOS at the lower gap edge and a
corresponding power law at the upper gap edge. These characteristic features
however vanish rapidly with increasing $U$, and already for $U=3$ the DOS 
mainly consists of
the Hubbard peaks at $\omega=+ U/2$ 
and $\omega=- U/2$ for $\sigma = \downarrow$ and $\sigma = \uparrow$,
respectively; reminiscent of the behavior expected for the Mott-Hubbard
insulator, where only the incoherent charge excitation peaks at high energies
are present \cite{mit,review_dmft,jtp92,bul99}. Note that neither from the
spectra in Fig.~\ref{fig:DOShalf} nor from the behaviour of the magnetic
moment in the inset of Fig.~\ref{fig:DOShalf} one can infer that 
at $U\approx 4.1$ the Mott-Hubbard metal-insulator transition occurs in the
paramagnetic state \cite{jtp92,bul99}.

\begin{figure}[htb]
\begin{center}
\mbox{}
\psfig{figure=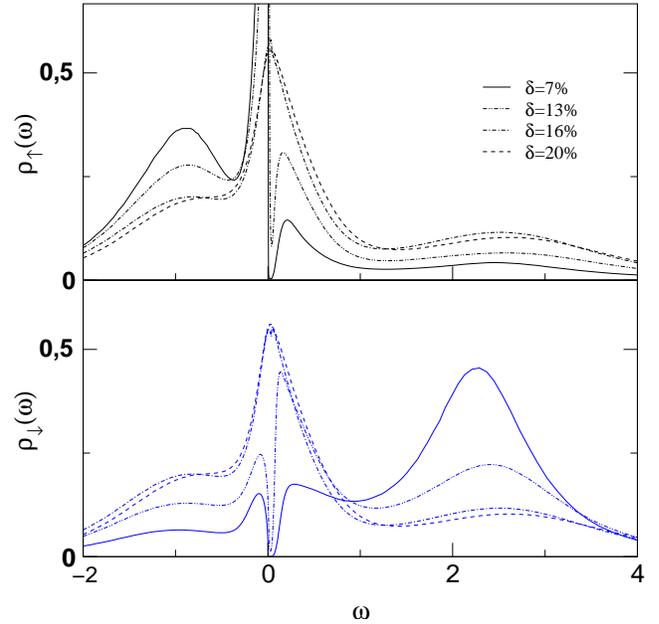,width=0.47\textwidth}
\end{center}
\caption{DOS for spin up and down for $U=3$ and different dopings $\delta=7\%$,
$\delta=13\%$, $\delta=16\%$ and $\delta=20\%$. The system
 at $\delta=20\%$ is already in
the paramagnetic phase.\label{fig:DOSofn}}
\end{figure}
Keeping $U$ fixed at $U=3$ and increasing $\delta$ leads to the spectra shown
in Fig.~\ref{fig:DOSofn}. Quite interestingly, the typical weak-coupling
characteristics reappear in the spectra for small doping and are still recognizable
for $\delta=13\%$. Note also that upon variation of doping and hence
of the magnetization 
the spectra are not shifted in the same way as in Hartree theory.
Instead, the dominant effect is a strong redistribution of spectral
weight from the Hubbard bands to the Fermi level. Eventually,
in the paramagnetic phase one recovers the well-known three peak structure
of the doped Hubbard model in the DMFT \cite{review_dmft}.

\begin{figure}[htb]
\begin{center}
\mbox{}
\psfig{figure=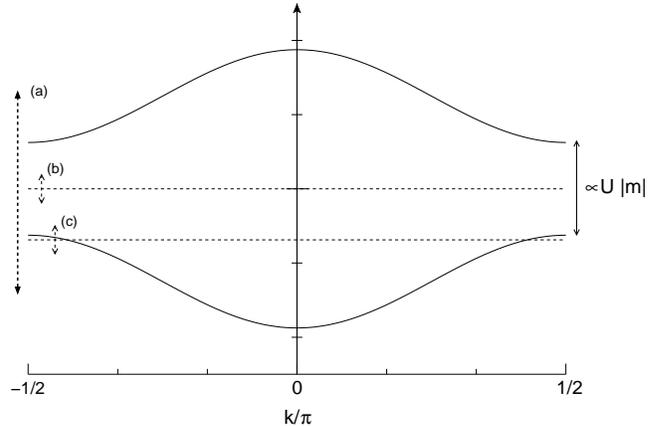,width=0.47\textwidth}
\end{center}
\caption{Schematic picture of the Hartree bandstructure of the Hubbard model
in the N\'eel state. The arrows at the left hand side of the figure
represent the energy scales of the corresponding {\em paramagnetic}  Fermi liquid
for half filling and weak
coupling (a), half filling and intermediate coupling (b) and finite doping and
intermediate coupling (c).\label{fig:eofk_afm}}
\end{figure}

The evolution of the spectra both at and off half filling can be understood
within a simple picture. In Fig.~\ref{fig:eofk_afm} we show a sketch of
the Hartree bandstructure of the Hubbard model in the N\'eel state, which has
two branches in the MBZ and a gap of width $\propto U|m|$ between them.
If, on the other hand, we inspect the {\em paramagnetic} solution, one for example finds
at half filling and for small values of $U$
a Fermi liquid with quasiparticles
defined on an energy scale larger than $U|m|$. This situation is indicated
by the arrow labeled (a) on the left side of Fig.~\ref{fig:eofk_afm}. Here
we expect, and indeed find for the
antiferromagnetic solution (see full curve in Fig.~\ref{fig:DOShalf}), a DOS that
shows the characteristic van-Hove singularities of Fig.~\ref{fig:DOSH}.
Increasing $U$ eventually leads to a situation, where the energy scale for
the quasiparticles in the paramagnetic state 
is finite but much smaller than $U|m|$ (arrow (b) in
Fig~\ref{fig:eofk_afm}). The self-energy in the energy region of the van-Hove
singularities then has a large imaginary part and will completely smear out
the characteristic structures. Further increasing $U$ into the Mott-Hubbard
insulator will then not change the picture qualitatively, explaining the
similarity between the curves for $U=3<U_{\rm MIT}$ and $U=6>U_{\rm MIT}$ in
Fig.~\ref{fig:DOShalf}. With finite doping, we move the chemical potential
 into, e.g., the lower band; this means that even for a relatively small quasiparticle
energy scale one again sees the van-Hove singularities at the band edge, which
results in the well defined singularities in the spectra for small doping in
Fig.~\ref{fig:DOSofn}.

From the  occupation numbers $n_\sigma$ obtained after convergence
of the DMFT calculation one can 
\begin{figure}[htb]
\begin{center}
\mbox{}
\psfig{figure=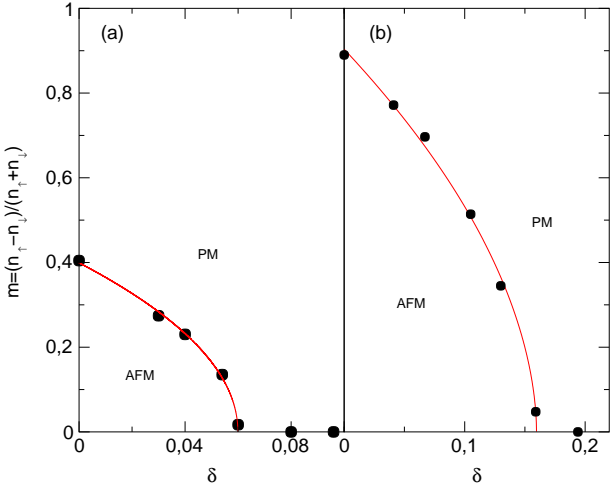,width=0.45\textwidth}
\end{center}
\caption{Doping dependence of the magnetization per electron for $U=1$ (a) and $U=3$ (b).
The full lines are fits with the function (\ref{equ:mofd}), the resulting
fit parameters are summarized in Table \ref{tab:fits}.%
\label{fig:mofd}}
\end{figure}
calculate the magnetization per electron,
$m=(n_\uparrow-n_\downarrow)/(n_\uparrow+n_\downarrow)$, as function of the
doping $\delta$. The results for $U=1$ and $U=3$ are shown in
Fig.~\ref{fig:mofd}a and b together with a fit to a power law
\begin{equation}
  \label{equ:mofd}
  m(\delta)=m_0\left|1-\frac{\delta}{\delta_c^{\rm AF}}\right|^\nu\;\;.
\end{equation}
The resulting fit parameters are summarized in Table~\ref{tab:fits}. As
expected for a mean-field theory, the  value for the critical
exponent is $\nu=1/2$. 

\begin{figure}[htb]
\begin{center}
\mbox{}
\psfig{figure=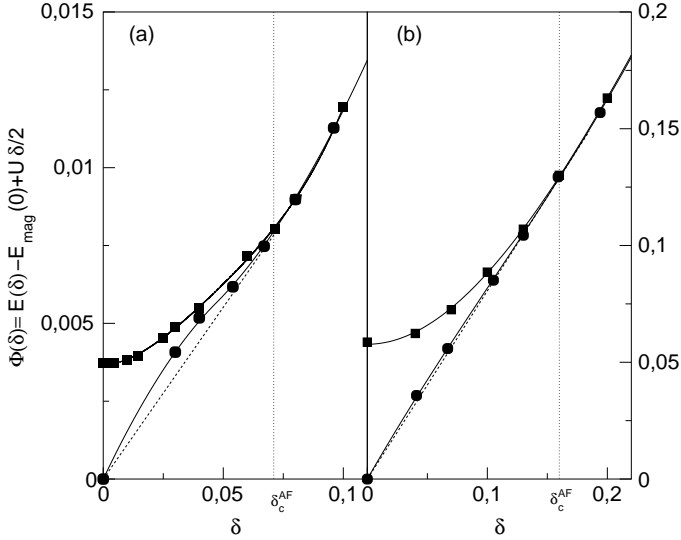,width=0.5\textwidth}
\end{center}
\caption{Doping dependence of the energy of the paramagnetic phase (squares)
and the N\'eel phase (circles) for  $U=1$ (a) and $U=3$ (b).
The full lines are fits with the function (\ref{equ:Eofd}), the corresponding
fit parameters are summarized in Table \ref{tab:fits}. The dashed lines are the
result of a Maxwell construction for the ground state energy.\label{fig:ps}}
\end{figure}
Finally, with the converged DMFT self-energy $\Sigma_\sigma(z)$ we can calculate
the expectation value $\langle H\rangle/N$ according to Eqs.~(\ref{equ:E}) and
(\ref{equ:Htpm}) respectively (\ref{equ:Htafm}) for the paramagnetic and
antiferromagnetic phase. The results for the characteristic function
$$
\Phi(\delta)=E(\delta)+\frac{U}{2}\delta - E_{\rm mag}(0)
$$
are summarized in Fig.~\ref{fig:ps}a and
b. In Fig.~\ref{fig:ps} the energies of the antiferromagnetic phase are
represented by the circles, those of the paramagnetic phase by squares.
The full lines interpolating the antiferromagnetic data are fits to
the function
\begin{equation}
  \label{equ:Eofd}
  \Phi(\delta)=\alpha\Phi^{\rm H}(\delta/\delta_1)
+\gamma\left(\frac{\delta}{\delta_1}\right)^3
\end{equation}
with $\Phi^{\rm H}(x)$ according to (\ref{equ:Phi}). The fit parameters are summarized
in Table~\ref{tab:fits}. The use of the function $\Phi^{\rm H}(x)$ in
\begin{table}[ht]
\begin{center}
\begin{tabular}{c||c|c|c||c|c|c|c}
\multicolumn{1}{c}{} & \multicolumn{3}{c}{Magnetization} & \multicolumn{3}{c}{Energy}\\
$U$ & $m_0$ & $\delta_c^{\rm AF}$ &$ \nu$ & $\delta_c^{\rm PS}$ & $\delta_1$ & $\alpha/\alpha^{\rm H}$
& $\gamma$\\
\hline
$1$ & $0.4$ & $0.06$ & $0.49$ & $0.07$ & $0.047$ & $0.52$ & $0$\\
$3$ & $0.9$ & $0.16$ & $0.54$ & $0.157$ & $0.191$ & $0.33$ & $0.026$\\
\end{tabular}
\end{center}
\caption{Results of the fits of $m(\delta)$ in Fig.~\ref{fig:mofd} to
expression (\ref{equ:mofd}) and $E(\delta)$ in Fig.~\ref{fig:ps} to
(\ref{equ:Eofd}).\label{tab:fits}}
\end{table}
(\ref{equ:Eofd}) is motivated by the results of van Dongen \cite{PvD_ps}.
The lines interpolating the paramagnetic data are meant as guides to the eye only. The dotted vertical lines
denote the value $\delta_c^{\rm AF}$ as obtained from Fig.~\ref{fig:mofd}.

The antiferromagnet obviously has the lower energy as compared to the paramagnet
in the region $0\le\delta\le\delta_c^{\rm AF}$. However, in both cases $U=1$
and $U=3$ we find a clear non-convex behavior in $E(\delta)$ in that region,
i.e.\ the aforementioned signature of an instability towards phase separation.
The true ground state energy as function of $\delta$ is obtained again via
a Maxwell construction, leading to the dashed lines in Fig.~\ref{fig:ps} and
the values $\delta_c^{\rm PS}$ given in Table~\ref{tab:fits}. Note that
in both cases $\delta_c^{\rm AF}\approx\delta_c^{\rm PS}$ within the
accuracy of the fitting procedure. 

While for $U=1$ the function $\Phi(\delta)$ nicely follows the weak-coupling
prediction (\ref{equ:Egwk}) with renormalized constant $\alpha$ one finds
a sizeable contribution $\sim\delta^3$ for $U=3$. This additional term 
results in a much weaker non-convex behavior of $E(\delta)$ for $U=3$.

For values $U>4$ we were not able to find a stable solution with N\'eel order
and well-defined doping $\delta>0$, although for 
$\delta<\delta_c(U)$ the paramagnetic phase becomes unstable. 
However, the numerical calculations rather produce a cycle 
encompassing a range of fillings instead of one solution with definite
filling here. It might be interesting to note that at least each of the 
fillings in this cycle has a unique magnetization associated with it and that
all spectra in this cycle correspond to an insulator.
Currently it is neither clear what type of magnetic solution we find here, nor
whether the breakdown of the N\'eel solution is a true physical effect or due
to numerical problems. Since at half filling the N\'eel state is present
at these values of $U$, incommensurate structures or again a phase separated
state seem to be possible.

For values of $U$ beyond $U_c\approx25$ yet another magnetic phase appears,
namely the ferromagnet. The existence of this phase has been observed in the
case of a hypercubic lattice and $U=\infty$ \cite{obermeier} and for a
generalized fcc lattice \cite{ulmke} before. Since these calculations had to be
done at finite and comparatively high temperatures, questions regarding the
ground state magnetization and, especially in the case of a hypercubic lattice,
the actual extent of the ferromagnetic phase in $(\delta,U)$ space could not
be discussed satisfactorily.

\begin{figure}[htb]
\begin{center}
\mbox{}
\psfig{figure=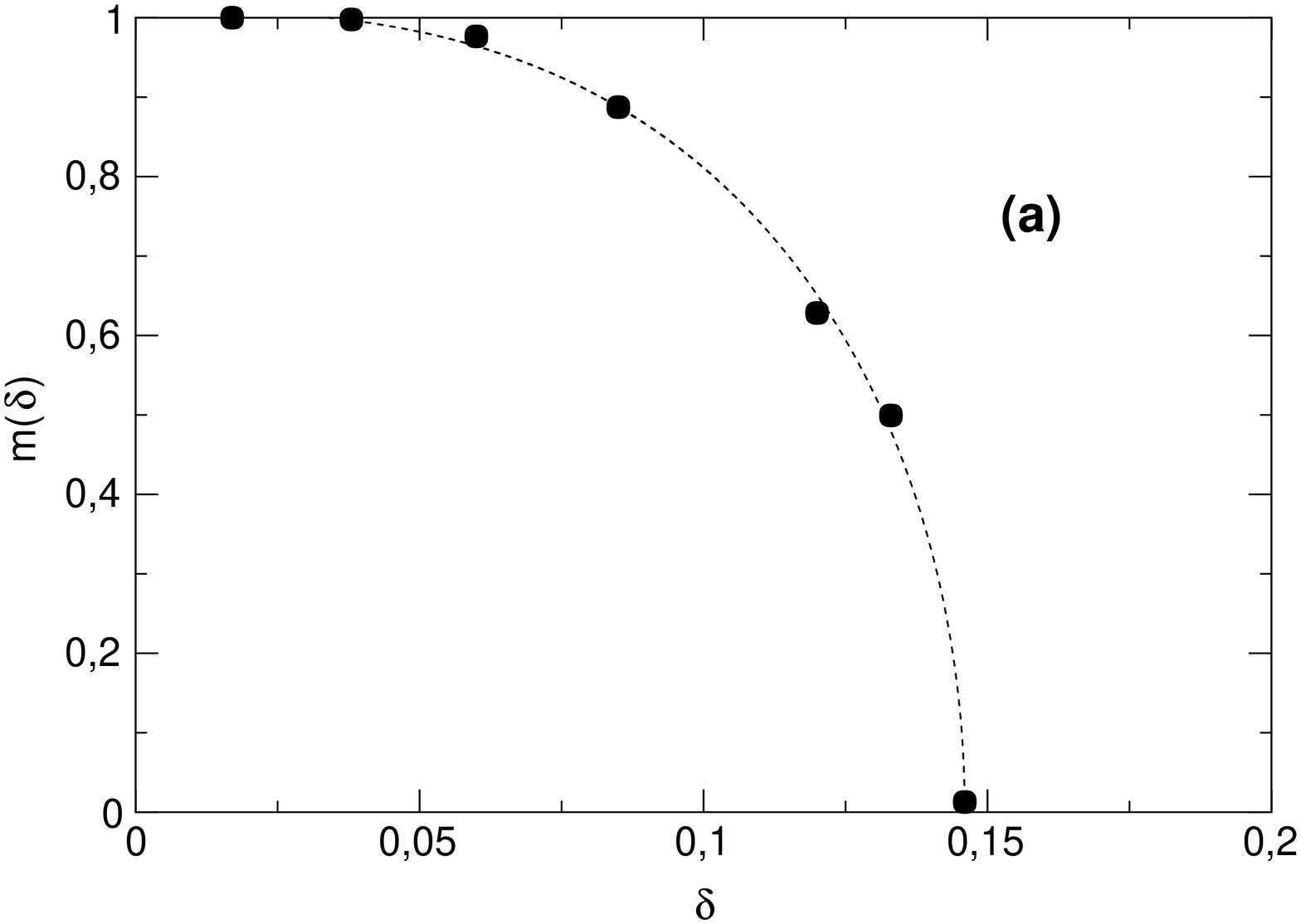,width=0.45\textwidth}
\mbox{}\\
\mbox{}
\psfig{figure=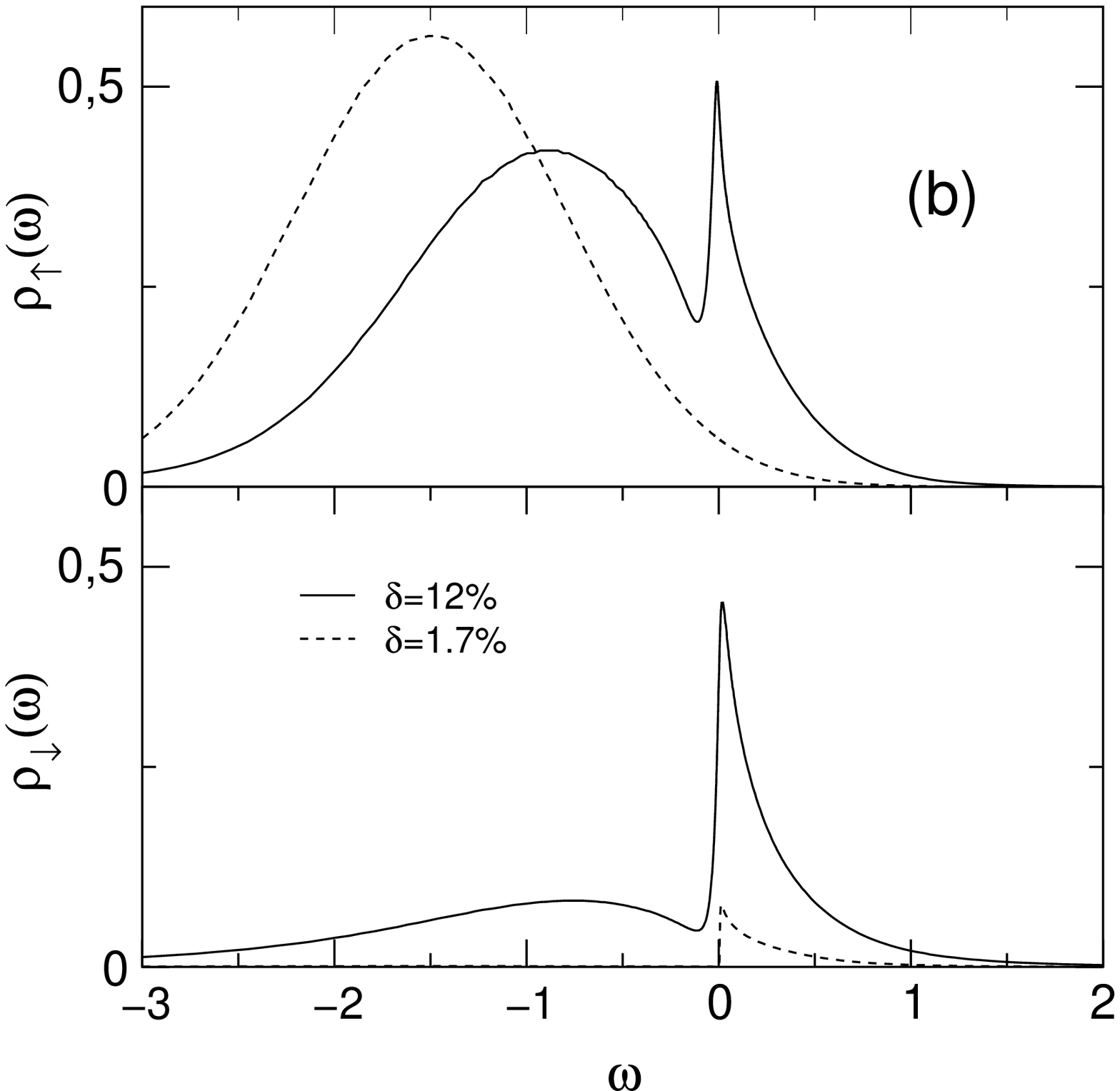,width=0.45\textwidth}
\mbox{}\\
\end{center}
\caption{(a) Ferromagnetic magnetization {\em per electron}\/ as function of
doping $\delta$ for $U=50$.
The full line is a fit to the function (\ref{equ:fitfm}). The critical doping
is $\delta_c\approx14.6$\%. Note that for $\delta\to0$ 
the results are consistent with a fully polarized ferromagnetic state.
(b) Local density of states for $U=50$ and two
characteristic dopings $\delta=12$\% (full lines) and $\delta=1.7$\% (dashed
lines). In contrast to the Stoner theory, one finds comparatively small shifts
in the spectra, but a strong redistribution of spectral weight.
\label{fig:fm}}
\end{figure}

As an example for the ferromagnetic phase at $T=0$ Fig.~\ref{fig:fm} shows the 
magnetization per electron, $m(\delta)=
(n_\uparrow-n_\downarrow)/(n_\uparrow+n_\downarrow)$
as function of doping (Fig.~\ref{fig:fm}a) and the
local DOS for two dopings (Fig.~\ref{fig:fm}b)
($U=50$). The data for $m(\delta)$ in Fig.~\ref{fig:fm}a
are fitted to the function
\begin{equation}\label{equ:fitfm}
m(\delta)=m_0\cdot\sqrt{1-\left(\frac{\delta}{\delta_c}\right)^\nu}\;\;,
\end{equation}
and the result is given by the dotted line. The parameters for the
fit are $m_0=1$, $\delta_c=14.6$\% and $\nu=2.75$. While for $\delta\nearrow\delta_c$
the typical mean-field behavior, i.e.\ $m(\delta)\propto\sqrt{1-\frac{\delta}{\delta_c}}$, is obtained, the result for
$\delta\to0$ is rather unconventional, viz $m(\delta)\propto1-\frac{1}{2}
\left(\frac{\delta}{\delta_c}\right)^{2.75}$. This fit assumes that a fully
polarized state is only reached as $\delta\to0$ \cite{naga_stab}. Note, however,
 that the numerical results
for the magnetization $m(\delta)$ for small $\delta$ are also consistent with a
fully polarized ferromagnet at finite $\delta$. 

It is also quite apparent from
the DOS in Fig.~\ref{fig:fm}b, that the ferromagnetism found here cannot be
understood on the basis of the typical Stoner theory. In contrast to the shifts of the
spectrum expected in the latter, we observe a strong redistribution of spectral
weight instead, but retain otherwise the typical structures due to
the strong correlations. Only in the case $\delta\to0$ the spectrum again
resembles that of a free system for the (almost completely polarized)
majority spins. The minority spins become strongly depleted below the Fermi
energy, the spectral weight can be found almost completely in the upper Hubbard
band situated around $\omega\approx U/2$ (not shown in the figure). Nevertheless,
we observe a tiny resonance {\em just above}\/ the Fermi energy even as $\delta\to0$.

\section{Summary and conclusion\label{sec:summary}}

In this paper, we used the dynamical
mean-field theory together with Wilson's
numerical renormalization group to investigate the ground-state
properties of the Hubbard model on a hypercubic lattice with nearest-neighbor
hopping both at and off half-filling. While at half-filling the ground-state
is antiferromagnetic for all $U>0$, at least for the weak and intermediate
coupling regime this magnetic order can only be realized
in a phase-separated state for any finite doping, thus supporting and extending
earlier weak-coupling predictions.

The mapping of the Hubbard model for large $U$ to an antiferromagnetic
$t$-$J$ model strongly suggests the dominance of antiferromagnetism in the
ground state.
The results for the Hubbard model in this paper show, however, 
that the type of magnetic order for  intermediate values of
the Coulomb repulsion $U$ off half filling is still an open issue; furthermore,
the role of phase-separation (which is observed for
$U\le 3$) has still to be clarified for larger values of $U$.

\begin{figure}[htb]
\begin{center}
\mbox{}
\psfig{figure=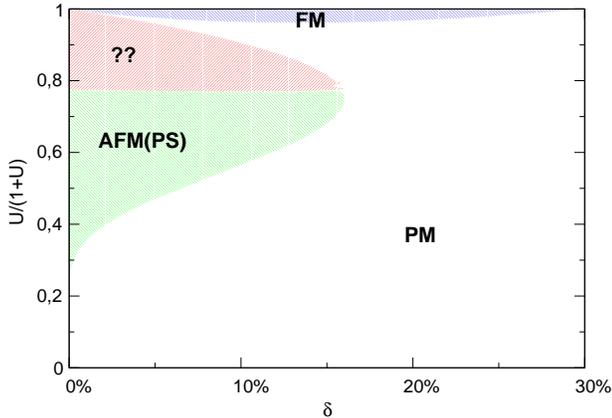,width=0.45\textwidth}
\end{center}
\caption{Schematic ground state phase diagram. At half-filling  ($\delta=0$),
the ground state is antiferromagnetic.
Close to half filling
and small $U$, a phase separated antiferromagnet is realized.
For values of $U$ beyond $U\approx4$ a magnetic phase is observed, whose
precise character could however not be identified.
For very large $U>25$ and up to $\delta\approx30$\% a ferromagnet is
found.\label{fig:hmpd}}
\end{figure}

The results are summarized in the schematic $(\delta,U)$ ground state phase diagram of
Fig.~\ref{fig:hmpd}.
To allow the inclusion of all values $0\le U<\infty$, the ratio $U/(1+U)$ is
used on the abcissa. Close to half filling  we find a phase separated N\'eel
antiferromagnet up to a certain value of $U<4$.
 The magnetization
as function of doping follows a typical mean-field behavior in all cases studied
and the spectra show the characteristic van-Hove singularities of the
band structure in the N\'eel state in cases where the characteristic
energy scale of the paramagnet is large enough. Most important is the
observation that, as typical for correlated ordered systems, the spectra
are not strongly shifted, as e.g.\ predicted by Hartree theory, but rather
show a strong redistribution of spectral weight. 

For values of $U>4$ the system
shows the tendency towards a magnetic instability, which could not be further
identified due to technical problems in the solution of the DMFT self-consistency.
However, we at least can exclude ferromagnetism here and a speculative possibility
would be the occurence of incommensurate phases or magnetic phases with
additional charge order. While the former cannot be addressed easily within the
present method, the latter possibility will be investigated further.

At very large values of $U>25$, there is a region of ferromagnetism, extending
between $0<\delta<30$\% as $U\to\infty$ \cite{obermeier}.
For a fixed value of $U$, the magnetization per electron in the
ferromagnetic state shows the tendency to saturate; from the numerical data
it is of course impossible to reliably conclude whether the system is fully
polarized at a finite $\delta$ already or only as $\delta\to0$. The data
are consistent with both scenarios, but the latter is supported by analytical
treatments of the case $\delta\to0$. As in the case of the antiferromagnet,
the spectrum shows a rather strong redistribution of spectral weight, not
simply a shift of the features, as would be expected from
Stoner theory.

The phase diagram shows a peculiarity, which has alread been pointed out
by Obermeier et al.\ \cite{obermeier}. In the region of very large $U$ and
$\delta\to0$ there exists the possibility of a direct transition between
the ``antiferromagnetic'' phase  and the ferromagnet. As at the point $(\delta,U)=(0,\infty)$ all possible spin configurations are degenerate, one can speculate
how the phase diagram looks like as $(\delta,U)\to(0,\infty)$
\cite{priv_com}. Generic
\begin{figure}[htb]
\begin{center}
\mbox{}
\psfig{figure=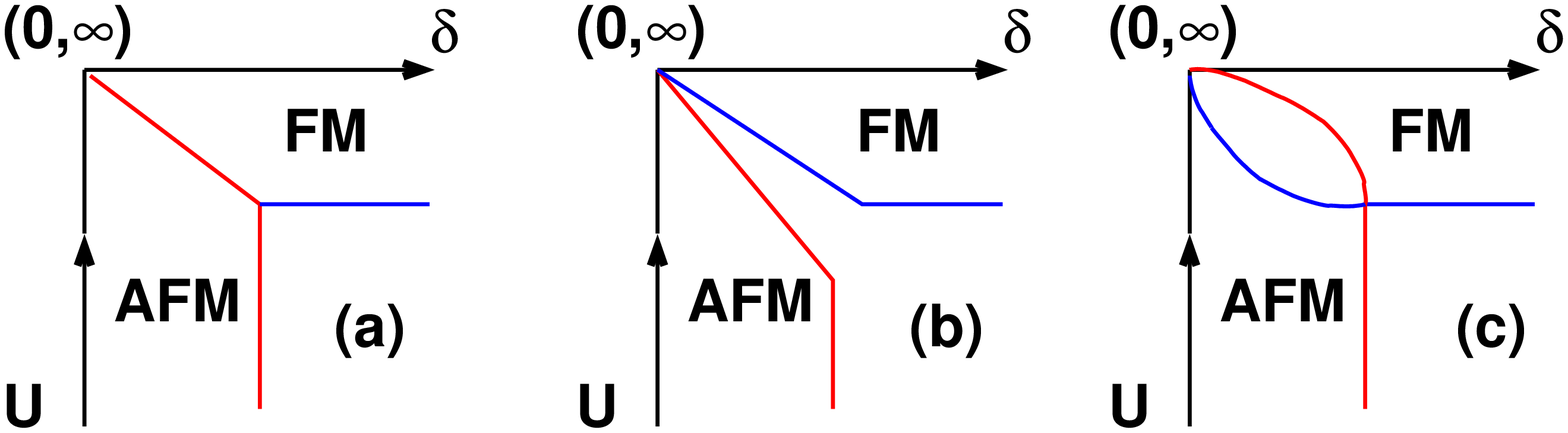,width=0.4\textwidth}
\end{center}
\caption{Possible realizations of the phase diagram as $(\delta,U)\to(0,\infty)$:
A direct transition between an antiferromagnet and a ferromagnet as in (a),
a small paramagnetic phase between the two las in (b) or a mixed type of 
phase (e.g.\ ferrimagnet) as in (c).\label{fig:SPEC}}
\end{figure}
possibilities are sketched in Fig.~\ref{fig:SPEC}. There can either be a direct
transition between the two phases (Fig.~\ref{fig:SPEC}a), which quite likely
would then be of first order, a gap filled by a
paramagnetic phase (Fig.~\ref{fig:SPEC}b) or a new phase, e.g.\ a ferrimagnet 
interpolating between the two extremes. The a priori exclusion or verification
of any of these structures is, without a detailed knowledge of the analytic
behaviour of the relevant quantities as function of $(\delta,U)$ in the vicinity of
$(\delta,U)=(0,\infty)$, not possible.

While there is a consensus about the magnetic properties of the Hubbard
model in a qualitative sense, the direct inspection of details still reveals
unexpected surprises. Even within the DMFT, where one can safely state that
the paramagnetic phase diagram including the Mott-Hubbard metal insulator
transition is now understood, the investigation of the magnetic properties
is far from complete. Obvious open questions are the magnetic properties at
intermediate values of $U$ and the behavior when the antiferromagnetic and
ferromagnetic phases meet. Furthermore, the behaviour of the magnetic phases,
especially the spectral properties in the ordered phases, in the presence
of a frustration due to longer range hopping, has not been addressed yet. This
might be of some interest regarding the question how the first order
Mott-Hubbard transition manifests itself in the magnetically ordered state.
Work along these lines is in progress.

\begin{acknowledgement}
We acknowledge useful discussions with
P.G.J.~van~Dongen,
M.~Jarrell,
J.~Keller,
D.~Logan,
M.~Vojta 
and
D.~Vollhardt.
This work was in part supported by the SFB 484 {\em Kooperative Ph\"ano\-mene im Festk\"orper} and the Leibniz Computer center.
\end{acknowledgement}

\end{document}